\documentclass[conference]{IEEEtran}
\IEEEoverridecommandlockouts
\usepackage{cite}
\usepackage{amsmath,amssymb,amsfonts}
\usepackage{algorithmic}
\usepackage{graphicx}
\usepackage{textcomp}
\usepackage{xcolor}
\usepackage{hyperref}
\usepackage{tikz}
\usepackage{flushend}

\def\BibTeX{{\rm B\kern-.05em{\sc i\kern-.025em b}\kern-.08em
    T\kern-.1667em\lower.7ex\hbox{E}\kern-.125emX}}

\IEEEaftertitletext{\vspace{-1\baselineskip}}

\newcommand*\circled[1]{\tikz[baseline=(char.base)]{
		\node[shape=circle,draw,inner sep=0.75pt] (char) {\textbf{#1}};}}
	
\begin{document}

\title{\scalebox{0.85}{Learning to Boost the Efficiency of Modern Code Review}}

\author{\IEEEauthorblockN{Robert Heumüller}
\IEEEauthorblockA{Otto von Guericke University Magdeburg, Germany \\
robert.heumueller@ovgu.de}
}

\maketitle

\begin{abstract}
		Modern Code Review (MCR) is a standard in all kinds of organizations that develop software.
MCR pays for itself through perceived and proven benefits in quality assurance and knowledge transfer.
However, the time invest in MCR is generally substantial.
The goal of this thesis is to boost the efficiency of MCR by developing AI techniques that can partially replace or assist human reviewers.
The envisioned techniques distinguish from existing MCR-related AI models in that we interpret these challenges as graph-learning problems.
This should allow us to use state-of-science algorithms from that domain to learn coding and reviewing standards directly from existing projects.
The required training data will be mined from online repositories and the experiments will be designed to use standard, quantitative evaluation metrics.
This research proposal defines the motivation, research-questions, and solution components for the thesis, and gives an overview of the relevant related work.
\end{abstract}


\section{Motivation\label{motivation}}Today, many organizations - ranging from open-source projects to global players like Microsoft or Google - have adopted some variation of a lightweight yet systematic peer code review process \cite{10.1145/3183519.3183525,10.5555/2486788.2486882}.
To distinguish them from previous more rigid practices, contemporary processes that are tailored to the needs of specific development teams and projects are known as Modern Code Review (MCR) \cite{10.1145/2491411.2491444}.
For example, the MCR process at Google involves five steps: \emph{Creating, Previewing Changes, Commenting, Addressing Feedback, and Approving Changes} \cite{10.1145/3183519.3183525}.

To get an idea of the average time invested for MCR, we interviewed project leads and developer teams of three medium-sized projects (\( >25 \) kSLOC) at a mid-sized software-development company in Germany\footnote{The author has been working part-time as a product owner and as a systems architect at that company for several years.}. 
In addition to these qualitative assessments, we also consulted statistics of the project management software, to determine how much time issues spend in the code-review state.
The results indicate that, between projects, 5.2h-10h per week (13\%-25\%) of developers' time is spent on MCR at that company.
These results are similar to what was reported for open-source projects (6.4h/wk) but higher than the findings of a case study at Google (3.2h/wk)\cite{Bosu2013,10.1145/3183519.3183525}.
While these numbers' generality is limited, we believe they do illustrate the potential gains of optimizing MCR processes.
\textbf{Thus, the goal of this thesis is to design AI models capable of assisting or partially replacing human reviewers, to reduce the time per review without compromising review quality. Trained on source code and review data, these models should learn both coding and reviewing standards from real-world data.}
\section{Research Questions\label{rqs}}Using only source code, review comments and meta-data:

\subsection*{RQ1: How can we design a model that can partially replace a human reviewer by predicting some aspects of their comments with respect to a given code change?}
\begin{figure}[t]
	\includegraphics[width=0.95\linewidth]{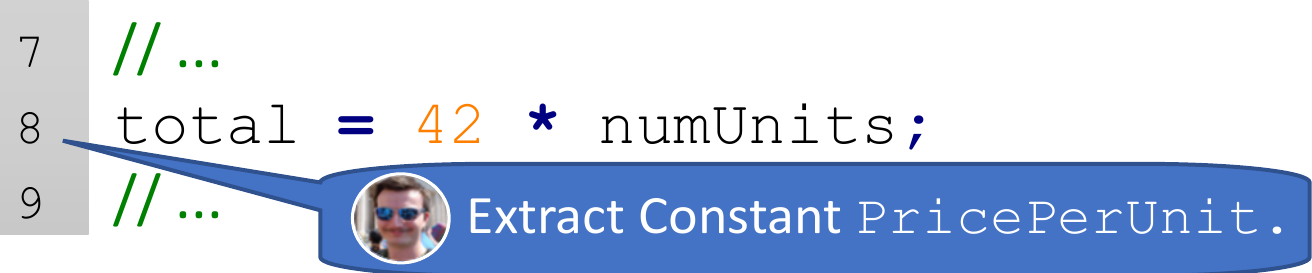}
	\caption{Example MCR comment requesting a refactoring\label{fig-cr-example}}
\end{figure}
\begin{figure}
	\includegraphics[width=0.85\linewidth]{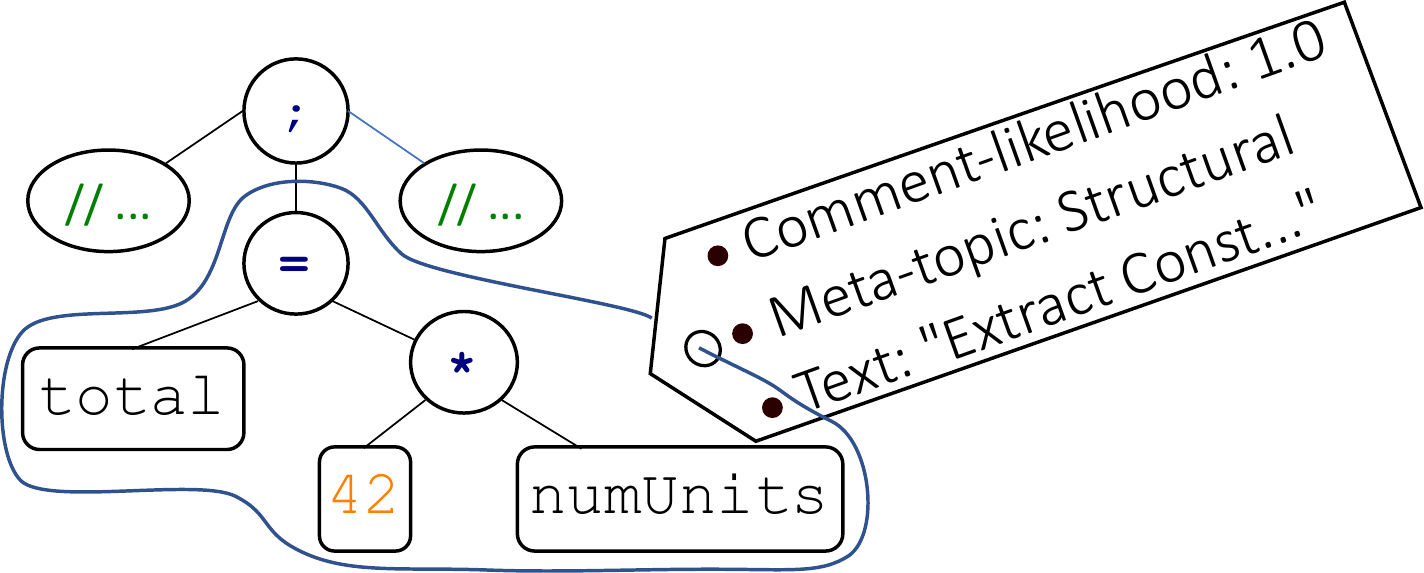}
	\caption{Labeled AST for the example from Figure \ref{fig-cr-example}\label{fig-ast-label}}
\end{figure}

Here, the goal is to predict different aspects of what a human reviewer would remark on in an MCR situation.
This includes predicting the comment locations, i.e. likelihood scores for the line numbers of a change under review, predicting the comment \emph{meta-topics} (e.g. style-violations, structural issues, bugs, use-case issues) for these locations, and ultimately generating meaningful comment texts.

The first major challenge is deciding what can be learned only from source code and review data, to accordingly define the concrete learning tasks.
For example, some comments address issues between an implementation and its software specification. 
Since a formal specification is typically not available for most source code, predicting such comments will be impossible in the majority of cases.
To overcome such problems, we are already working on models classifying the \emph{meta-topic} of given comments.
This should enable us to define achievable learning tasks, for example detecting style-violations, structural issues, or some types of bugs.

The second challenge is providing the necessary ground-truth, including positive and negative labels for features like the comment-likelihood. How do you identify code that is not likely to be commented on?
Our starting point is to use code that was recently changed due to a review but then remained stable for a minimum period.

The third challenge is designing and evaluating AI models for the defined learning tasks.
Here, the different learning tasks induce particular types of models such as regression, classification, or text generation.
However, we interpret all learning tasks as graph-learning problems for program representation graphs, e.g. ASTs (Abstract Syntax Trees).
All models will thus learn to compute features, to classify, or to generate text for graphs or subsets of their nodes and edges. 
Figure \ref{fig-cr-example} illustrates the graph-learning idea for ASTs using the source code and review comment from Figure \ref{fig-ast-label}.

\subsection*{RQ2: How can we design and train a model that judges the quality of review comments concerning a given change?}
Here, the goal is to learn to quantify the quality of code review comments, e.g. for providing feedback to reviewers.
The definition of \emph{quality} encompasses many aspects, some of which were explored in related work (cf. Section \ref{stateofart}).
As a starting point, we define two features, \emph{actionability}, i.e. whether the comment induces changes to the respective code, and \emph{clarity}, e.g. how much further discussion the comment induced.
Formally, we then interpret the judging of comment quality as a multimodal embedding and regression problem.

\subsection*{RQ3: How can the methods developed in RQ1 and RQ2 be effectively integrated into a real-world MCR workflow?}
Here, the goal is to develop a practical assistance architecture that integrates the models from \emph{RQ1} and \emph{RQ2} to provide valuable assistance in different phases of MCR processes.
We outline two examples for Google's review flow (cf. \ref{motivation}): In the \emph{previewing changes} phase, likely comment-locations and topics could be highlighted, and in the \emph{commenting phase}, reviewers would receive feedback on comment usefulness.
We anticipate that such assistance should help to increase MCR efficiency, which we intend to evaluate in an empirical study.
Another important consideration for this RQ is how to keep models up to date, for example by utilizing online training.
\section{Solution Components\label{roadmap}}We propose a concept involving four major components.

\circled{1} Dataset: First, the essential data of software and code reviews must be sufficient in quality, quantity, and diversity. 
Thus, we will mine large-scale online repositories, for example specialized review tools like Gerrit, and, in particular, GitHub pull-requests.
We are currently working on a first dataset for \emph{Elasticsearch} that will include comments (\(\approx \! 100k\)), their corresponding change hunks, and relevant java file revisions (\(\approx\!47k\)).
Datasets must then be curated and labeled with input- and ground-truth-features, e.g. comment meta-topics.

\circled{2} Input Representation: Second, we will analyze suitable input representations for source code, review comments, and meta-data. 
We will focus on graph-based program representations, i.e. ASTs, or other specialized representations, for two reasons:
First, it has been observed that several program-analysis tasks benefit from structure-aware representations \cite{8668037,10.1145/3290353}.
Second, it allows us to experiment with recent graph-learning algorithms, which achieve state-of-the-art performance for various graph-analysis tasks \cite{Chen2019b,Kipf2017,Velickovic2017}.
Similarly, for review comments, numerous synthetic representations and representation learning techniques can be explored \cite{Jones1973,Sutskever2014,Kim2014,Vaswani2017,Mikolov2013b}.

\circled{3} Model Design and Evaluation: Third, we want to design AI models that can learn relationships between source code and review data.
Following the standard procedure, we intend to split the dataset into training and testing sets.
For \emph{RQ1} we will then train models to predict, e.g., whether the AST of a particular change will be commented on or what such a comment's meta-topic will likely be.
Example models for \emph{RQ2} will either process only comments, or jointly process comments and ASTs, to estimate how helpful a comment may be, first generally, and then respecting a particular change.
For all evaluations we aim to compute standard accuracy metrics on the separate testing data to detect over-fitting.

\circled{4} Practical Evaluation: Fourth, we will implement relevant parts of the architecture from \emph{RQ4}, probably as plugins to existing MCR tools. Then, we want to perform a case-study with professional developers to assess the real-world impact on efficiency and to gain insights into developer acceptance.
\section{Related Work\label{stateofart}}Many studies have analyzed aspects of the effectiveness of MCR \cite{10.5555/2819009.2819015,10.1145/2884781.2884840,10.1145/3387904.3389270,McIntosh2015,7081827,7332454,10.1109/TSE.2009.27,10.1145/2597073.2597076,Kononenko2015,Edmundson2013,Paixao2019,7180075}.
However, proving MCRs effectiveness is not our research focus.
Instead, for \emph{RQ2} we are interested in what metrics they defined to quantify MCR effectiveness, e.g. good code-reviews uncover bugs or improve the software design.

An important distinction to previous work on code-review analysis is that our envisioned methodology will learn to directly correlate source code and review data. 
To the best of our knowledge, this has not been successfully attempted before.
The most similar, yet still conceptually different, related approach analyzed how reviews can be mapped between projects via code clone detection \cite{8668037}.

Some review datasets which we can build on have been previously published  \cite{8595176,Yang2016,Mukadam2013}.
However, after our preliminary screening, we believe that further repository mining is necessary to meet the particular needs of our research.
Particularly \emph{CROP}\cite{8595176} gives important insights into the challenges and pitfalls of constructing review datasets.

Regarding neural and graph-based program-representations, we can draw on many promising, recent approaches \cite{Zhang2019,Mou2014,Chen2019a,Wei2017,10.1145/3290353}.
Further, various general graph-learning algorithms could be adapted, e.g. graph-convolutional- and graph-attention-networks \cite{Kipf2017,Velickovic2017}. 
A recent survey is also available\cite{Chen2019b}. 

For representing comments, many existing learning techniques for text could be relevant, particularly word embeddings \cite{Mikolov2013b}, RNN- \cite{Sutskever2014}, CNN-\cite{Kim2014}, and attention-models \cite{Vaswani2017}.

For both representation learning tasks, we are particularly interested in attention-based approaches \cite{Vaswani2017} for their superior trainability and ability to focus on relevant sub-structures of graphs and long sequences.

\bibliographystyle{IEEEtran}
\bibliography{IEEEabrv,thesis_proposal}  
% Generated by IEEEtran.bst, version: 1.14 (2015/08/26)

\end{document}